\pdfoutput=1
\documentclass[aps,prd,twocolumn,nofootinbib,superscriptaddress]{revtex4-2}

\usepackage{graphicx}
\usepackage{amsmath}
\usepackage{amssymb}
\usepackage[utf8]{inputenc} 
\usepackage{url} 

\begin{document}
	
	
\title{Is Bianchi I a Bouncing Cosmology in the Wheeler-DeWitt picture?}
	
	
\author{Eleonora Giovannetti}
\email{eleonora.giovannetti@uniroma1.it}
\affiliation{Dipartimento di Fisica (VEF), P.le A. Moro 5 (00185) Roma, Italy}
\author{Giovanni Montani}
\email{giovanni.montani@enea.it}
\affiliation{ENEA, Fusion and Nuclear Safety Department, C.R. Frascati, Via E. Fermi 45 (00044) Frascati (RM), Italy}
\affiliation{Dipartimento di Fisica (VEF), P.le A. Moro 5 (00185) Roma, Italy}
	
	
\date{\today}
	
\begin{abstract}
We provide a quantum picture for the emergence of a bouncing cosmology, according to the idea that a semiclassical behavior of the Universe towards the singularity is not available in many relevant Minisuperspace models. In particular, we study the Bianchi I model in vacuum adopting the isotropic Misner variable as an internal clock for the quantum evolution. The isomorphism between the Wheeler-DeWitt equation in this Minisuperspace representation and the Klein-Gordon one for a relativistic scalar field allows to identify the positive and negative frequency solutions as associated to the collapsing and expanding Universe respectively. We clarify how any Bianchi I localized wave packet unavoidably spreads when the singularity is approached and therefore the semiclassical description of the model evolution in the Planckian region loses its predictability. Then, we calculate the transition amplitude that a collapsing Universe is turned into an expanding one, according to the standard techniques of relativistic quantum mechanics, thanks to the introduction of an ekpyrotic-like matter component which mimics a time-dependent potential term and breaks the frequency separation. In particular, the transition probability of this \textquotedblleft Quantum Big Bounce\textquotedblright acquires a maximum value when the mean values of the momenta conjugate to the anisotropies in the collapsing Universe are close enough to the corresponding mean values in the expanding one, depending on the variances of the in-going and out-going Universe wave packets. This symmetry between the pre-Bounce and post-Bounce mean values reflects what happens in the semiclassical bouncing cosmology, with the difference that here the connection of the two branches takes place on a pure probabilistic level.

\end{abstract}


\maketitle

\section{Introduction \label{I}}
The presence of an initial singularity in the Universe thermal history \cite{LL,PC} constitutes the most relevant shortcoming of the implementation of General Relativity to the cosmological problem. Since the Seventies, the idea of a possible Bounce was formulated to replace the initial singularity and reconnect a collapsing Universe to our expanding one, in order to depict the scenario of a cyclical Universe \cite{W}. Many implementations of a Big Bounce scenario have been considered over the years, essentially based on suitable modifications of the Einstein theory of gravity (for recent examples, see\cite{B1,B2,O}). 

However, the justification of a bouncing cosmology as the result of a quantum gravity effect in the Planckian epoch arose when in \cite{RS} it was demonstrated that the kinematical spectrum of the geometrical operators possesses a discrete nature in the framework of Loop Quantum Gravity (LQG). Indeed, the implementation of this formulation \cite{ashtekar2003,ashtekar2005gravity,Ashtekar2006,AshtekarI,Ashtekar2008,Ashtekar2011, Bojowald2002,Bojowald2004} to the cosmological problem provided the emergence of a Big Bounce with a minimal Universe volume in the past being different from zero and, consequently, a regularized behavior of the energy density 
(for critical considerations on the 
so-called Loop Quantum Cosmology (LQC) see \cite{C1,C2,Bojowald2020}). The presence of a similar behavior of the Universe can be also recovered when Polymer Quantum Mechanics (PQM) \cite{Pol} is applied to the cosmological degrees of freedom \cite{M,Ant,EFG,S} (for a comprehensive review on the bouncing cosmologies in PQM and LQC see \cite{review}).

All these descriptions of 
a bouncing cosmology from quantum physics mainly rely on the characterization of 
quasi-classical states for the Universe, which outline a mean behavior as following a Big Bounce picture and so deviating from General Relativity at sufficiently high energy density. However, this description seems lacking when the considered quasi-classical state follow the Bounce trajectory with a significant spreading that would bring the dynamics into a pure
quantum sector. Actually, it cannot be excluded that, during the evolution of a contracting isotropic Universe, relevant anisotropies can arise, see for instance the so-called ekpyrotic Universes \cite{ekp,ekp2,shear} where such a problem is addressed. The emergence of non-negligible anisotropy degrees of freedom constitutes a crucial mechanism, through which a closed Universe would pass from a Robertson-Walker geometry to a Bianchi IX one \cite{PC}. In this picture, the Kasner-like behavior of a Bianchi IX localized state would be continuously perturbed towards the Bounce by the potential term, see for instance the numerical analysis in \cite{F}. 

Based on these considerations, it appears more reasonable to consider the Bounce as a pure quantum region of the Universe evolution and so apply to its description the concept of probability amplitude, associated to a transition from a collapsing Universe to an expanding one. In this respect, see the pioneering work \cite{HH} for approaches based on a path integral formulation applied to the quantum cosmology in the Euclidean sector, while for recent applications to the Lorentz sector see \cite{L,LDT}. In this framework, if we consider the Big Bounce as the transition amplitude to pass with a non-zero probability from a collapsing Universe to an expanding one, the need of a semiclassical minimal value of the Universe volume is no longer essential in order to deal with a bouncing cosmology. 

The aim of this analysis is investigating the possibility to have a Quantum Big Bounce also in the Wheeler-DeWitt (WDW) approach of quantum cosmology. We study the metric canonical quantization of the Bianchi I model in vacuum, adopting the well-known Misner variables \cite{Misner69,PC}. According to the standard literature \cite{DeWitt67}, by choosing the isotropic Misner variable as the internal time, we are able to provide an isomorphism between the WDW equation and a massless Klein-Gordon one. Comparing the behavior of the classical constants of motion with their corresponding quantum eigenvalues, we can interpret the negative and positive frequencies of the WDW solutions as states which describe the expanding and collapsing Universe respectively. Also, we theoretically and numerically show that the localized wave packets are subjected to a significant spreading process, in order to support the need of describing the behavior of the Bianchi I model towards the singularity as an intrinsic quantum phenomenon. 

Then, we include a matter term with an equation of state parameter $w>1$ that breaks down the frequency separation, being a time-dependent potential responsible for the transition from a 
collapsing Universe to an expanding one (i.e. the  positive and negative frequency states). We remark that the standard theory of relativistic scattering processes are used, as discussed in \cite{BjD}, where the projection of in-going states onto out-going ones is described via the wave function formalism (here the WDW wave function of the Universe is used), so that we escape the so-called third quantization of the 
cosmological field and all the ambiguous related issues \cite{K,BjD}. In particular, we project the in-going wave packet for the collapsing Universe, that represents the exact solution of the WDW equation during the ekpyrotic phase, onto a Bianchi I expanding 
wave packet, according to the procedure presented in \cite{BjD}. In both the Universe wave packets, we use a Gaussian weight in the momenta with non-zero mean values. As a result of treating the Quantum Big Bounce as a scattering process, the probability amplitude of transition from the collapse to the expansion is non-zero and mathematically well-defined. In particular, the probability amplitude has a peaked profile with the interpretation that the most likely transition takes place when the mean value of the momenta 
of the expanding wave packet is approximately equal to the mean values of the contracting one, depending on their variances. By other words, a localized collapsing state of the Universe has the maximum probability to make the transition into an expanding localized state, if the morphology of the latter closely resembles the packet shape of the former.

This result opens a new perspective 
on the physical nature of the Big Bounce, at least when an internal time variable can be properly recovered. Indeed, the possibility for a quantum transition in the canonical quantum dynamics phenomenologically appears as a bouncing cosmology; nevertheless, it is due to the mixing of positive and negative frequency solutions when an interaction term is included, and does not rely on the existence of a minimal semiclassical value of the Universe volume.

The paper is structured as follows. In Sec. \ref{Min} the Minisuperspace of the Bianchi models is introduced, with a particular focus on the dynamics of a Bianchi I wave packet in the subsection \ref{BI}. In Sec. \ref{BDs} the procedure of scattering integrals using the wave function formalism in the Klein-Gordon theory is presented. Sec. \ref{S} contains the core of the work. The transition amplitude from a contracting to an expanding Bianchi I Universe thanks to the $w>1$-matter term is developed and the existence of a Quantum Big Bounce in the WDW theory from a probabilistic point of view is discussed. Finally, in Sec. \ref{C} some concluding remarks are outlined. We note that we use $8\pi G =c=\hslash=1$ throughout the article.


\section{Minisuperspace of the Bianchi models \label{Min}}

Let us start our analysis by discussing the structure of the Hamiltonian constraint of the Bianchi cosmological models, i.e. anisotropic, homogeneous and non-stationary Universes, in order to outline the isomorphic feature of the Minisuperspace with the relativistic quantum theory. In the following, we will concentrate our attention to the Bianchi I model, that one having zero spatial curvature. 

Using the Arnowitt-Deser-Misner (ADM) formalism and the Misner variables $(\alpha,\beta_+,\beta_-)$, the line element describing a Bianchi cosmology takes the form 
\cite{LL,PC,MBBI} 
\begin{equation}
	ds^2 = N^2dt^2 - e^{2\alpha}
	\left(e^{2\beta}\right)_{ab}\sigma^a\sigma^b
	\, , 
	\label{eletc1}
\end{equation}
where $\beta\equiv diag\{\beta_++\sqrt{3}\beta_-,\beta_+-\sqrt{3}\beta_-,-2\beta_+\}$, $N$ denotes the lapse function and all the variables are time-dependent only, due to the spatial homogeneity. The 1-forms $\sigma^a$ 
($a=1,2,3$) reflect the specific isometry of the considered Bianchi model and in the case of Bianchi I they reduce to exact differentials. 

The action describing the dynamical features of a Bianchi model takes the expression
\begin{equation}
	S_B = \int dt\,(
	p_{\alpha}\dot{\alpha} + p_+\dot{\beta}_+ + p_-\dot{\beta}_- - NH)\,,\label{eletc2} 
\end{equation}
where
\begin{equation}
	 H\equiv Ce^{-3\alpha}\left[-p_{\alpha}^2 + p_+^2 + p_-^2 + 
	 e^{4\alpha}V_B(\beta_+,\beta_-) + \lambda e^{-3(w-1)\alpha}\right]
	\,
	\label{eletc3}
\end{equation}
in which the explicit form of the potential term $V_B$ fixes the considered Bianchi model 
($V_B\equiv 0$ for Bianchi I).
Here, the dot symbol denotes the derivative with respect to $t$, $C$ is a constant depending on the performed spatial integration and $p_{\alpha}$, $p_+$ and $p_-$ are the respective conjugate momenta to the Misner variables. 
The isotropic variable $\alpha$ defines the Universe volume, while $\beta_+$ and $\beta_-$ are the real gravitational degrees of freedom since they correspond to the model anisotropies. Also, a matter content with equation of state $P=w\rho$ has been added in the action. In particular, we consider $w>1$, i.e. an ekpyrotic phase in the Bianchi I Universe, in order to deal with a dominant term near the singularity.  

As already outlined in \cite{DeWitt67} for the case of a generic Superspace, 
the variable associated to the volume has a different signature with respect to the gravitational degrees of freedom and therefore it can be interpreted as a time variable for the classical and quantum dynamics of the system.
By other words, we are entitled to 
adopt $\alpha$ as the internal clock of our Minisuperspace corresponding to the homogeneous cosmologies. It is worth expressing the link of the relational time $\alpha$ in terms of the generic time variable $t$ by varying the action \eqref{eletc2} with respect to the momentum $p_{\alpha}$, namely
\begin{equation}
	\dot{\alpha} = - 2NCe^{-3\alpha}p_{\alpha}
	\, . 
	\label{eletc4}
\end{equation}
If we choose the synchronous time ($N\equiv 1$), i.e. the time coordinate in which the thermal history of the Universe is preferably described, we see that for $p_{\alpha}<0$ the physical space expands with time, while for $p_{\alpha}>0$ it contracts as time goes. Furthermore, we note that the momentum $p_{\alpha}$ becomes a constant of motion if we deal with a Bianchi I model (for which $V_B\equiv 0$) and we require $\lambda \equiv 0$, so that its sign can be specified \emph{a priori} and the two branches of the expanding and collapsing Universe can be separated at a classical level. 

Clearly, when we canonically quantize the dynamical system described in \eqref{eletc2} and \eqref{eletc3} we lose the information contained in 
\eqref{eletc4}, since all the physical content is summarized in the Universe wave function $\psi=\psi(N,\alpha ,\beta_\pm)$ selected by the
Hamiltonian operator $\hat{H}$ that annihilates it. Indeed, the canonical implementation of the primary constraint $\hat{p}_N\equiv 0$ ($p_N$ being the conjugate momentum to $N$) provides that the wave function is independent of the lapse function, while the secondary constraint, whose classical existence is ensured by the variation of the action \eqref{eletc2} with respect to the lapse function $N$, reads as
\begin{equation}
	\hat{H}\psi = \left[ \Box + e^{4\alpha}V_B(\beta_+,\beta_-) + \lambda e^{-3(w-1)\alpha}\right] \psi (\alpha,\beta_\pm) = 0\,,
	\,
	\label{eletc5}
\end{equation}
where $\Box=\partial_{\alpha}^2 - \partial^2_{\beta_+} - \partial^2_{\beta_-}$. As we can see, the WDW equation written in the Misner variables still outlines the role of the volume-like coordinate $\alpha$ as the internal time of the system and the parallelism with a Klein-Gordon relativistic equation in the presence of a time-dependent potential is almost immediate. Also, it is easy to check that equation \eqref{eletc5} admits the probability density
\begin{equation}
	j_0 = i\left( \psi^*\partial_{\alpha} \psi - \psi\partial_{\alpha}\psi^*\right)
	\,
	\label{eletc6}
\end{equation}
in analogy with the Klein-Gordon formalism. In order to deal with a positive defined probability density $j_0$, we need to perform the so-called frequency separation, that is clearly impossible in the presence of a potential term depending on the time variable $\alpha$, like in Eq. \eqref{eletc5}. On the other hand, in the simplest case of a Bianchi I model ($V_B\equiv0$) without any matter content, the frequency separation is easily reached since the Universe wave function can be written in the plane wave basis as 
\begin{equation}
	\psi^{\pm}_{\omega_k}(\alpha,\beta_\pm) = 
	e^{\mp i\omega_k\alpha}e^{i(k_+\beta_+ + k_-\beta_-)}
	\, , 
	\label{eletc7}
\end{equation}
where $\omega_k \equiv \sqrt{k_+^2 + k_-^2}$. 

Now, if we apply the quantum operator $\hat{p}_{\alpha} = -i\partial_{\alpha}$ to the wave function $\psi^\pm_{\omega_k}$ we see that the positive frequency solution is an eigenstate with a negative eigenvalue, while the positive one is associated to the negative frequency state. Accordingly, we can interpret the positive frequency solutions as corresponding to states that describe an expanding Universe, vice-versa the negative frequency solutions are associated to a collapsing Universe. These considerations are supported by the Ehrenfest theorem that ensures that equation \eqref{eletc4} is verified by the corresponding quantum expectation values.

This interpretation of the frequency separation will allow to deal with a relativistic quantum approach to the analysis of the Bianchi I dynamics, as discussed in \cite{BjD}, that is based on the use of the Universe wave function instead of the third quantization procedure. 

\subsection{Wave packets behavior in the Bianchi I Minisuperspace \label{BI}}

Before proceeding on analyzing in details the Big Bounce as a quantum process, in this subsection we preliminary discuss the properties of the Bianchi I wave packet. We construct a superposition of the particular solutions of the form \eqref{eletc7} by means of a generic localizing function, in order to satisfy the requirement of describing a quasi-classical state for the Universe which is compatible with the frequency separation. For example, in the case of an expanding Universe we have 
\begin{equation}
	\psi (\alpha ,\beta_\pm) = \iint_{-\infty}^{+\infty}dk_+dk_-\,A(k_+,k_-)\psi ^+_{\omega _k}\,, 
	\label{eletc8}
\end{equation}
where $A(k_+,k_-)$ is commonly chosen as a Gaussian function fixed by the initial condition on the wave function at a given instant of time $\alpha = \alpha_0$. 

It is important to stress 
that the probability density $j_0$ is strictly positive for monochromatic plane waves only (see \cite{P1,P2}). Indeed, it can be easily seen that just the superposition of two different plane waves with positive energy-like eigenvalues leads to the emergence of regions in which $j_0$ assumes negative values. This fact should be taken under serious consideration when we deal with Universe wave packets in quantum cosmology, since it raises the question that $j_0$ (properly a charge density) could not be a good candidate for a well-defined probability density through which computing the expectation values of the quantum operators of the theory. In this work we try to go beyond the semiclassical approach to the dynamics near the singularity and so we overcome the problem of a well-defined probability density by resorting to a pure quantum approach. Moreover, even the behavior of the Bianchi I wave packet highlights the weakness of the semiclassical approach, as showed in the following. 

\begin{figure*}
\begin{minipage}[h]{3.4cm}
	\includegraphics[width=1.8\linewidth]{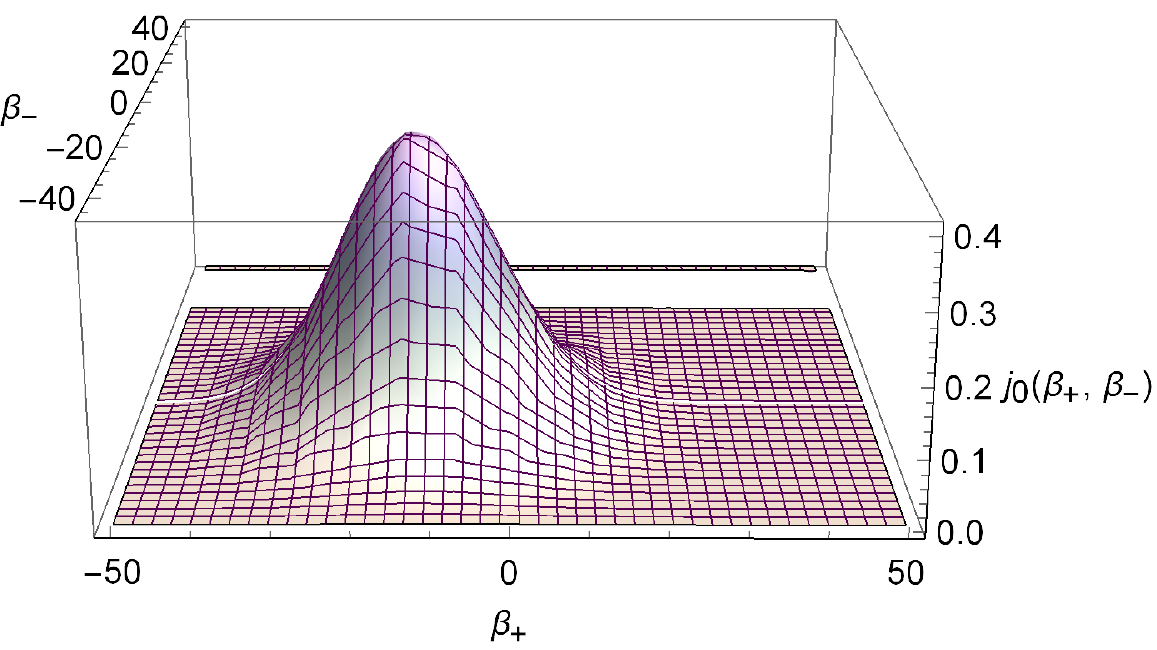}
\end{minipage}\qquad\qquad\qquad\qquad
\begin{minipage}[h]{3.4cm}
	\includegraphics[width=1.8\linewidth]{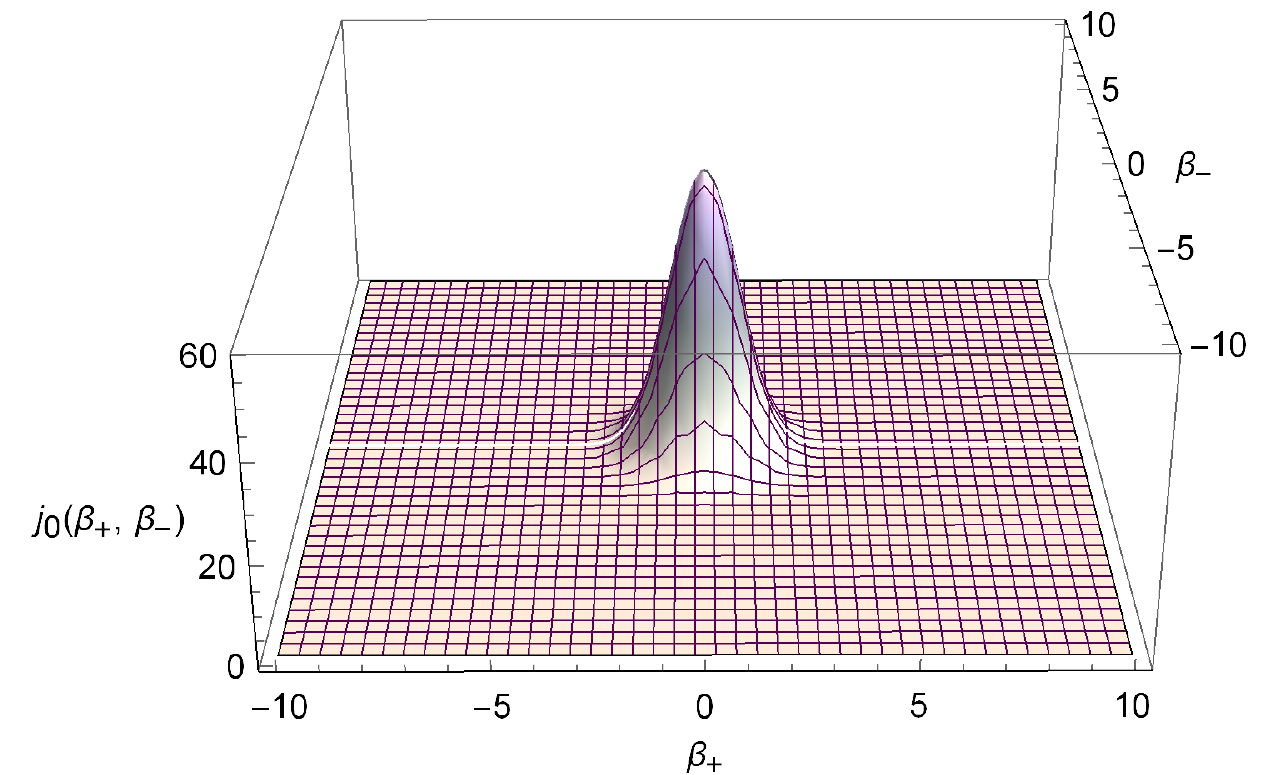}
\end{minipage}\qquad\qquad\qquad\qquad
\begin{minipage}[h]{3.4cm}
	\includegraphics[width=1.8\linewidth]{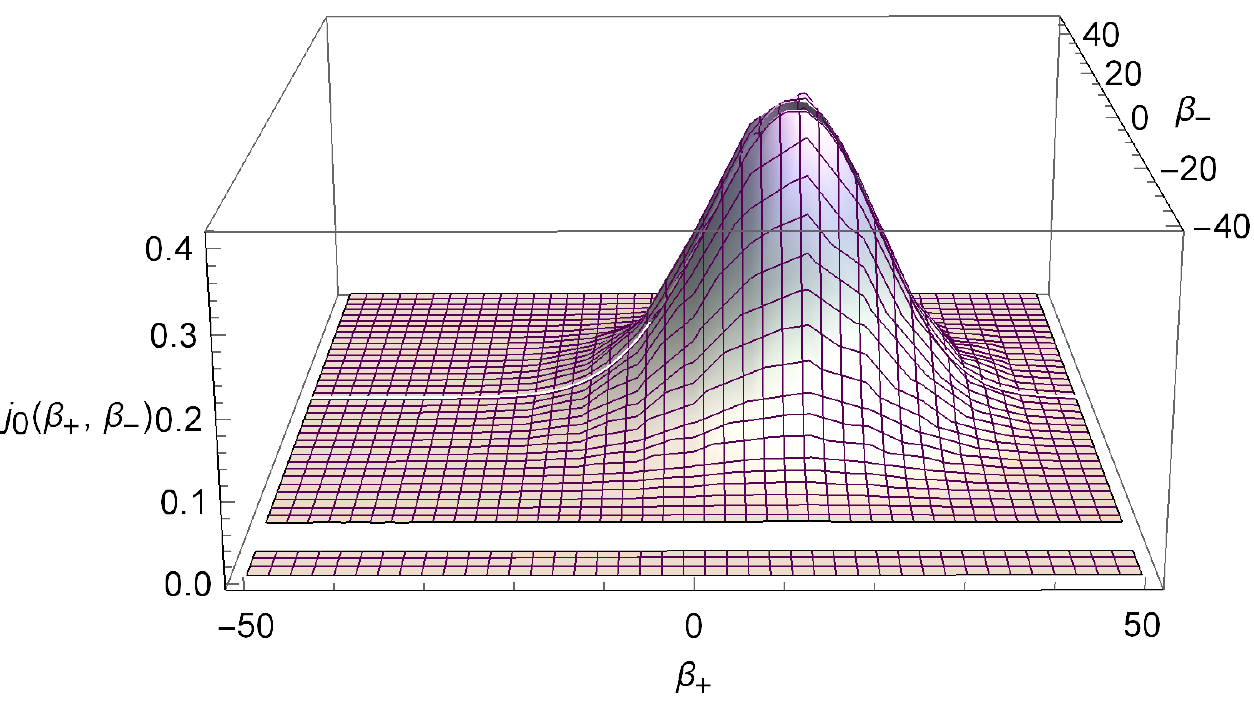}
\end{minipage}\qquad\qquad\qquad\qquad

\caption{3D-plots of the probability density $j_0$ associated to a Bianchi I wave packet containing only positive frequency plane waves. It is calculated at three different values of the relational time $\alpha$ ($\alpha=-10,0,10$ respectively).}
\label{j}
\end{figure*}

First of all, we notice that the Bianchi I wave packet \eqref{eletc8} is characterized by a non linear dispersion relation $\omega_k$. This feature produces a spreading of the wave packet during its propagation due to the presence of a non-zero second derivative of $\omega_k$ with respect to $k_+,k_-$. In fact, if we consider Gaussian coefficients of the form
\begin{equation}
\label{G}
A(k_+,k_-)=e^{\frac{-(k_+-\bar{k}_+)^2}{2\pi\sigma_+^2}}e^{\frac{-(k_--\bar{k}_-)^2}{2\pi\sigma_-^2}}
\end{equation}
in \eqref{eletc8}, we can reasonably suppose that these coefficients are significantly different from zero only in the neighborhood of $(\bar{k}_+,\bar{k}_-)$ and so justify an expansion of $\omega_k$ up to the second order term in $(k_+,k_-)$ that simplifies the analytical calculation of the integral. As it can be easily demonstrated, a linear term in $\alpha$ enters in the $\sigma$ of the Gaussian packet due to the second derivative of $\omega_k$ with respect to $k_+,k_-$. As the wave packet propagates, both the mean value and the variance change with time, producing the spreading phenomenon (see Fig. \ref{j}). This behavior does not affect the Friedmann-Lemaitre-Robertson-Walker (FLRW) Universe that is characterized by a linear dispersion relation. We notice that here the probability density $j_0$ remains always positive even if referred to a wave packet superposition, since the Gaussian weight privileges the peak frequency (see Fig. \ref{j}).
This feature of the Bianchi I Universe clearly prevents a satisfactory description of the dynamics towards the singularity by means of quantum expectation values on semiclassical states. More specifically, when more general cosmological models with respect to the FLRW one are considered, in view of producing a reasonable description of the Universe near the Planckian region, the hypothesis of a localized state is violated.

\section{The transition amplitude in the wave function formalism \label{BDs}}

In this section we describe the formalism at the basis of the scattering amplitude calculation by following the approach presented in \cite{BjD}. The Klein-Gordon equation describes relativistic particles of zero spin by means of its solutions, i.e. scalar wave functions. For the free particle the equation reads as
\begin{equation}
\label{KG}
(\Box+m^2)\varphi(x)=0\,,
\end{equation}
whose solution $\varphi(x)$ can be written as a superposition of plane waves with both positive and negative frequencies
\begin{equation}
	f^{(\pm)}_\mathbf{p}(x)=\frac{e^{\mp ip\cdot x}}{\sqrt{(2\pi)^32\omega_\mathbf{p}}}
\end{equation}
where $\omega_\mathbf{p}=p_0>0$ and $p^2=m^2$ according to the Einstein energy condition. They form a complete set and satisfy the following orthogonality and normalization conditions

\begin{align}
	\label{KGnorm}
	&\int d^3x\,f^{(\pm)*}_\mathbf{p'}(x)i\overleftrightarrow{\partial_0}f^{(\pm)}_\mathbf{p}(x)=\pm\delta^3(\mathbf{p}-\mathbf{p'})\,,\\
	&\int d^3x\,f^{(\pm)*}_\mathbf{p'}(x)i\overleftrightarrow{\partial_0}f^{(\mp)}_\mathbf{p}(x)=0\,.
\end{align}
The Feynman propagator for the Klein-Gordon
equation has the expression
\begin{equation}
	\Delta_F(x'-x)=\int\frac{d^4p}{(2\pi)^4}\,\frac{e^{-ip\cdot(x'-x)}}{p^2-m^2+i\epsilon}
\end{equation}
and solves the following equation
\begin{equation}
	\label{KGP}
	(\Box_{x'}+m^2)\Delta_F(x'-x)=-\delta^4(x'-x)\,.
\end{equation}
It propagates the positive-frequency parts of a generic superposition of solutions forward in time
and the negative-frequency ones backward in time by construction. We notice that the possibility of creation and annihilation of single spinless
particles (confirmed by experimental observations below the particle creation energy threshold) would require a many-particle theory in interaction as developed in the quantum field theory formalism. However, it is possible to extend the propagator approach to the study of these particles coupled to source terms added to the
right-hand side of \eqref{KG}. In particular,
when an interaction term is added in Eq. \eqref{KG} it becomes

\begin{equation}
	\label{KGV}
	(\Box+m^2+V(x))\phi(x)=0
\end{equation}
and the general integral has the following form
\begin{equation}	\phi(x)=\varphi(x)-\int d^4y\,\Delta_F(x-y)V(y)\phi(y)\,,
	\label{phi}
\end{equation} 
through which the solution of \eqref{KGV} can be evaluated to the desired accuracy by iteration. In \eqref{phi}, $\varphi(x)$ is a superposition of plane waves defined as
\begin{align}
\varphi(x)&=\varphi^{(+)}(x)\,+\,\varphi^{(-)}(x)=\\&=\int d^3p\,c_+(\mathbf{p})f^{(+)}_\mathbf{p}+\int d^3p\,c_-^*(\mathbf{p})f^{(-)}_\mathbf{p}\,.
\end{align}
Now we can compute the transition amplitude to a particle
state of given momentum $p'$ by projecting the scattered wave emerging from the interaction onto a normalized free wave of
momentum $p'$, so that the transition probability is then given by the absolute square of this amplitude.
In particular, for particles and anti-particles scattering we have
\begin{equation}
	S_{\mathbf{p'_+},\mathbf{p_+}}=\delta^3(\mathbf{p'_+}-\mathbf{p_+})-i\int d^4y\,f^{(+)*}_{\mathbf{p'_+}}(y)V(y)\phi(y)
\end{equation}
and
\begin{equation}
	S_{\mathbf{p'_-},\mathbf{p_-}}=\delta^3(\mathbf{p'_-}-\mathbf{p_-})-i\int d^4y\,f^{(-)*}_{\mathbf{p'_-}}(y)V(y)\phi(y)
\end{equation}
respectively, whereas for pair production and annihilation we have
\begin{equation}
	S_{\mathbf{p_+},\mathbf{p_-}}=-i\int d^4y\,f^{(+)*}_{\mathbf{p'_+}}(y)V(y)\phi(y)	\label{Smatrix}
	\end{equation}
and
\begin{equation}
	S_{\mathbf{p_-},\mathbf{p_+}}=-i\int d^4y\,f^{(-)*}_{\mathbf{p'_-}}(y)V(y)\phi(y)
\end{equation}
respectively. We notice that, under the hypothesis of a limited interaction region, $\phi(y)$ reduces to plane waves for $t\rightarrow-\infty$ and for $t\rightarrow+\infty$ it can be expanded in the plane waves basis with the $S$-matrix elements as the expansion coefficients. So, the conservation over time of the Klein-Gordon norm \eqref{KGnorm} guarantees that $\phi(y)$ can be normalized as plane waves and, if $\phi_\mathbf{p}(y)$ (namely the solution of \eqref{phi} that reduces to a plane wave of momentum $\mathbf{p}$ for $t\rightarrow-\infty$) form a complete set, the unitarity of the operator $S_{Bounce}$ is ensured. This legitimates $\mathcal{P}=|S_{Bounce}|^2$ as a well-defined probability density.

\section{The transition amplitude from a collapsing to an expanding Bianchi I Universe \label{S}}

In this section we present the core of the work. First, we recall that the Bianchi I model is characterized by a primordial singularity at a classical level that is not solved in the WDW approach, even when an ekpyrotic phase is considered in the Kasner dynamics. Our aim is to investigate the probability of having a quantum transition from a collapsing to an expanding Universe, in effort to treat the Big Bounce as a relativistic quantum interaction transposed to the primordial cosmology when a time-dependent interaction term is present. In this way, the Big Bang singularity would be solved even in the WDW formulation at a probabilistic level. All the background theory used in the following has been presented in the previous section and based on \cite{BjD}. 

First of all, we solve the WDW equation for the Bianchi I model with a ekpyrotic-like matter term
\begin{equation}
	\hat{H}\psi = \left[ \partial_{\alpha}^2 - \partial^2_{\beta_+} - \partial^2_{\beta_-} + \lambda e^{-3\varepsilon\alpha}\right] \psi (\alpha,\beta_\pm) = 0
	\,,
	\label{eletc9}
\end{equation}
that corresponds to \eqref{eletc5} with $V_B\equiv0$ and $\varepsilon=w-1>0$. In this model, the role of the ekpyrotic-like matter component is that of a time-dependent potential responsible for the quantum transition and able to mix the positive and negative frequency states. 

We search for a solution in $L^2(\mathbb{R})$, namely of the form
\begin{equation}
	\psi (\alpha,\beta_\pm) = \varphi(\alpha)e^{ik_+\beta_+}e^{ik_-\beta_-}\,, 
\end{equation}
so that \eqref{eletc9} reduces to the following equation for the variable $\alpha$
\begin{equation}
	\partial^2_\alpha\varphi(\alpha)+(\omega_k^2+\lambda e^{-3\varepsilon\alpha})\varphi(\alpha)=0\,, 
	\label{alpha}
\end{equation} 
with $\omega_k$ defined as in \eqref{eletc7}. The exact solution reads as
\begin{equation}
	\varphi(\alpha)=c_1\varphi^{(-)}(\alpha)+c_2\varphi^{(+)}(\alpha)\,,
\end{equation}
where $c_1$, $c_2$ are integration constants and
\begin{align}
	&\varphi^{(-)}(\alpha)=J_{-\frac{2i\omega_k}{3\varepsilon}}(2\sqrt{\lambda e^{-3\varepsilon\alpha}}/3\varepsilon)\Gamma(1-\frac{2i\omega_k}{3\varepsilon})\,,	\label{varphi1}\\
	&\varphi^{(+)}(\alpha)=J_{\frac{2i\omega_k}{3\varepsilon}}(2\sqrt{\lambda e^{-3\varepsilon\alpha}}/3\varepsilon)\Gamma(1+\frac{2i\omega_k}{3\varepsilon})\,.
	\label{varphi2}
\end{align}
In \eqref{varphi1}-\eqref{varphi2}, $J_{\nu}(x)$ indicates the Bessel function of the first kind and $\Gamma(x)$ the Euler Gamma function. So, the general solution is
\begin{equation}
	\label{unicol}
	\psi(\alpha,\beta_\pm)=\iint_{-\infty}^{+\infty}dk_+dk_-\,A(k_+,k_-)\varphi(\alpha)e^{ik_+\beta_+}e^{ik_-\beta_-}\,,
\end{equation}
where $A(k_+,k_-)$ is defined as in \eqref{G}. We note that $\varphi^{(+)}(\alpha)$ and $\varphi^{(-)}(\alpha)$ reduce to a plane waves for $\alpha\rightarrow+\infty$(except for a phase depending on $\varepsilon$). In other words, $\varphi(\alpha)$ has the right limit far from the singularity since it reduces to the free solution when the $w>1$-matter component becomes negligible. It is worth noting that the Bessel functions (corresponding for $\lambda\rightarrow0$ to negative and positive frequency states respectively) are equally weighed in the in-going wave packet by setting $c_1=c_2=1/\sqrt{2}$. In this way we do not privilege neither the collapsing neither the expanding configuration. 

Now we compute the Big Bounce transition amplitude by projecting the scattered wave emerging from the interaction onto a free Universe wave packet of the form
\begin{equation}
	\label{uniexp}
\chi(\alpha,\beta_\pm)=\iint_{-\infty}^{+\infty}dk_+'dk_-'\,A(k_+',k_-')e^{-i\omega_{k'}\alpha}e^{ik_+'\beta_+}e^{ik_-'\beta_-}\,,
\end{equation} 
that consists in a superposition of only free expanding plane waves by means of Gaussian coefficients as defined in \eqref{G}, where the prime symbol identifies the out-going wave packet. In particular, by using \eqref{Smatrix} we obtain
\begin{equation}
	\label{SB}
	S_{Bounce}=-i\iiint_{-\infty}^{+\infty}d\alpha\,d\beta_+ d\beta_-\,\chi^*(\alpha,\beta_\pm)V(\alpha)\psi(\alpha,\beta_\pm)\,,
\end{equation}
where $\chi(\alpha,\beta_\pm)$ represents the free expanding Universe wave packet \eqref{uniexp}, $V(\alpha)=\lambda e^{-3\varepsilon\alpha}$ and $\psi(\alpha,\beta_\pm)$ is the Universe wave packet \eqref{unicol} that emerges from the interaction. The analytical calculation of the integral in the anisotropies $\beta_+,\beta_-$ produces the two Dirac delta functions $\delta(k'_+-k_+)$ and $\delta(k'_--k_-)$ that make the integration over $k'_+,k'_-$ trivial. We remark that the integral over the configurational variables $\alpha,\beta_+,\beta_-$ can be exchanged with those ones over $k_+,k_-$ and $k'_+,k'_-$ implicitly contained in the wave packets $\chi(\alpha_,\beta_+,\beta_-)$ and $\psi(\alpha_,\beta_+,\beta_-)$, since the integration domains are independent of all the variables. Then, the remaining integral in the variables $\alpha,k_+,k_-$ has been computed by means of both analytical and numerical methods. Finally, the Big Bounce transition probability is obtained by making the absolute square of \eqref{SB} through which we get a four-variables scalar function

\begin{equation}
|S_{Bounce}(\bar{k}'_+,\bar{k}'_-,\bar{k}_+,\bar{k}_-)|^2
\end{equation}
that depends only on the mean values of the momenta conjugate to the anisotropies $\beta_+,\beta_-$ of the in-going and out-going states, as expected (in this calculation all the variances of the Gaussian coefficients have been set to $\sigma=1/\sqrt{2\pi}$). In particular, by fixing the quantum numbers $\bar{k}_+,\bar{k}_-$ of the in-going wave packet we obtain a two-variables function

\begin{equation}
\mathcal{P}(\bar{k}'_+,\bar{k}'_-)
\end{equation} 
describing the probability of a Big Bounce transition.  

\begin{figure}[h!]
		\includegraphics[width=1\linewidth]{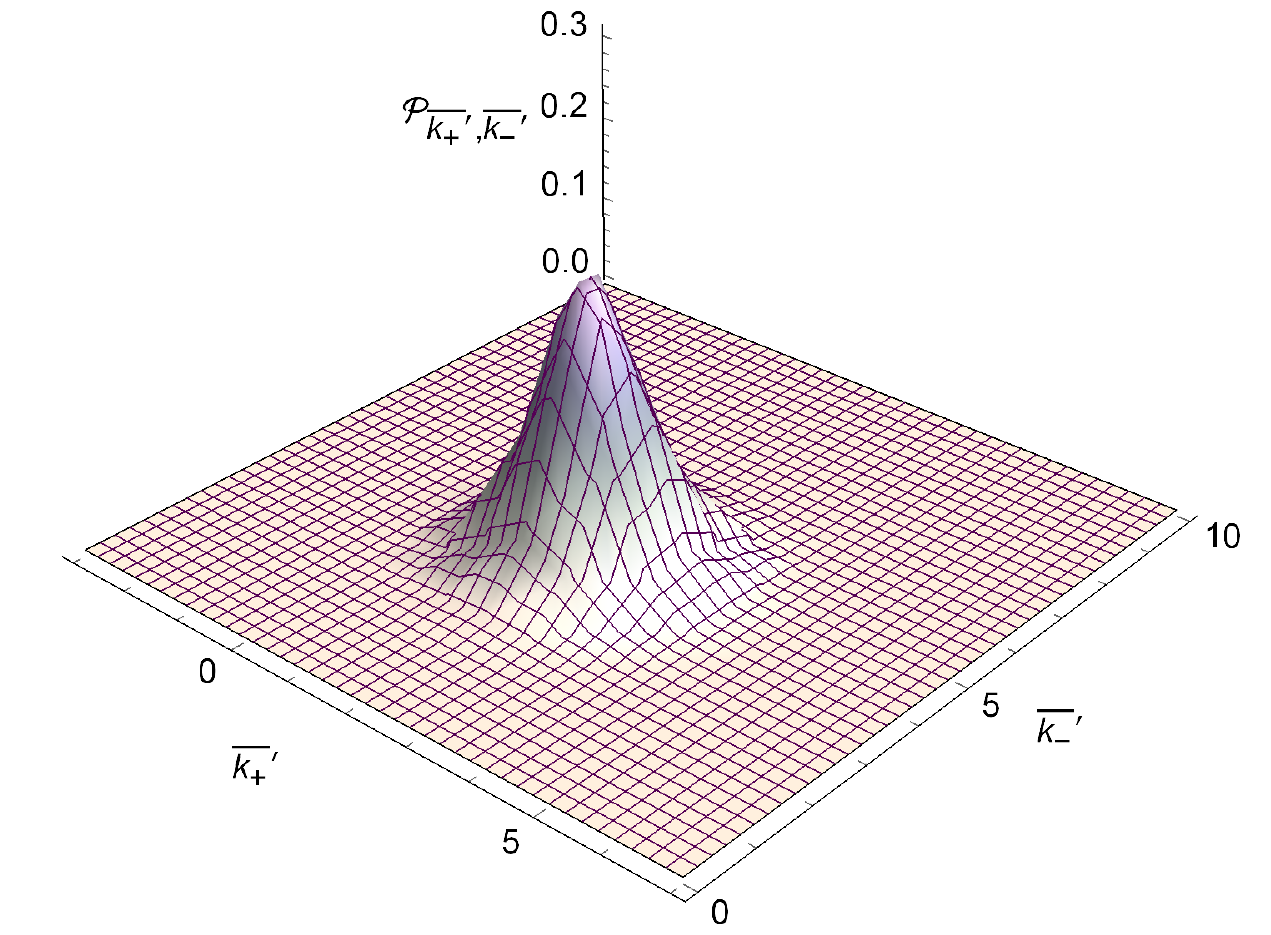}
		\includegraphics[width=0.85\linewidth]{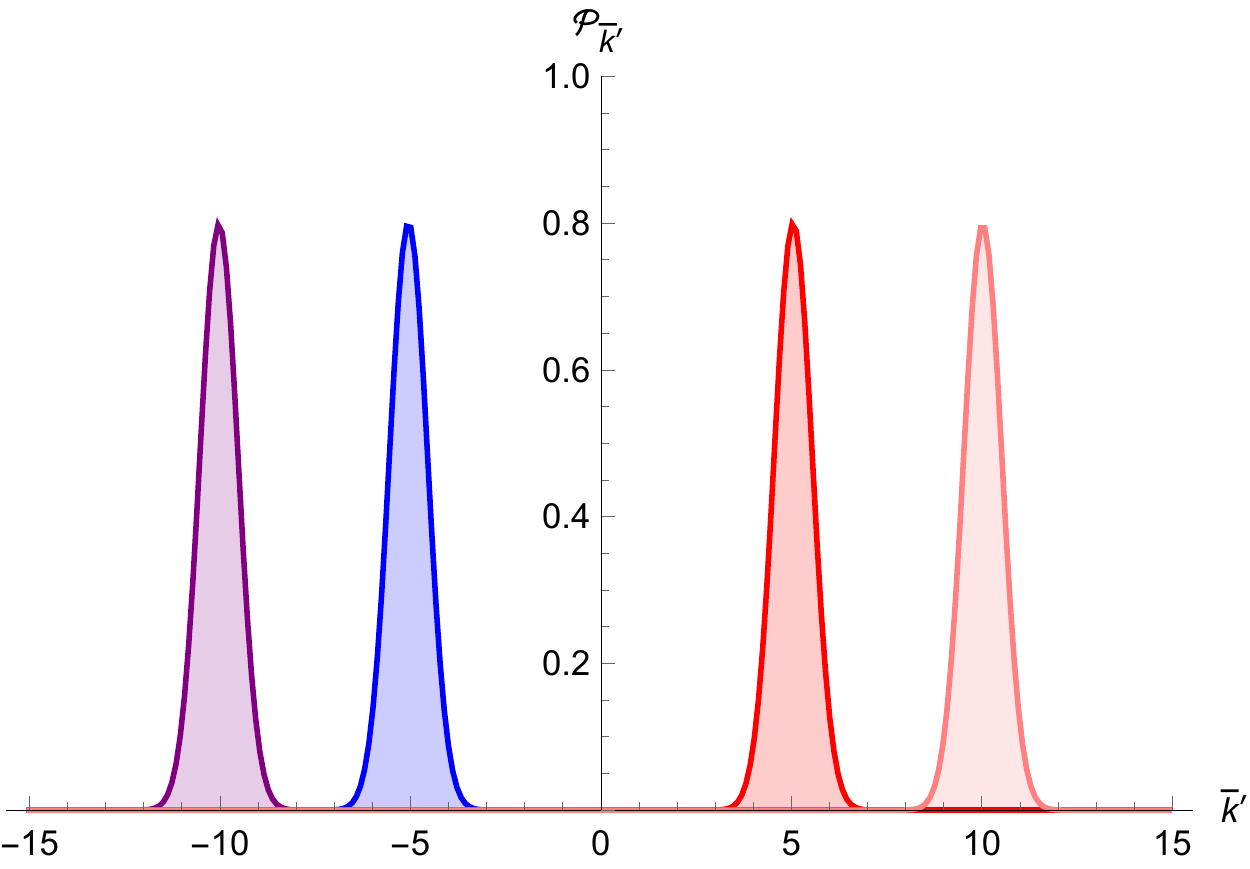}
	\caption{Top: 3D-plot of the normalized probability density $\mathcal{P}(\bar{k}'_+,\bar{k}'_-)$ of the Big Bounce transition. We can notice that the peak of $\mathcal{P}(\bar{k}'_+,\bar{k}'_-)$ occurs for $(\bar{k}'_+,\bar{k}'_-)=(\bar{k}_+,\bar{k}_-)$. In this graph we have considered $\lambda=1$, $\varepsilon=1/3$, $\bar{k}_+=2$, $\bar{k}_-=4$. Bottom: 2D-plots of the normalized probability density $\mathcal{P}(\bar{k}')$ of the Big Bounce transition for different values of $\bar{k}$. We can notice that the peak of $\mathcal{P}(\bar{k}')$ occurs for $\bar{k}'=\bar{k}$. In this graph we have considered $\lambda=1$, $\varepsilon=1/3$ and (from the left) $\bar{k}=-10,-5,5,10$.}
\label{prob}
\end{figure}

In Fig. \ref{prob} we present two plots that highlight the Gaussian shape of $\mathcal{P}(\bar{k}'_+,\bar{k}'_-)$, noticing that the peak of the probability density in the mean values $\bar{k}_+'$, $\bar{k}_-'$ of the out-going wave packet occurs always in correspondence of the values assigned to the corresponding mean values $\bar{k}_+$, $\bar{k}_-$ of the in-going wave packet. In particular, in the 3D-plot we can see that the peak occurs for  $(\bar{k}'_+,\bar{k}'_-)=(\bar{k}_+,\bar{k}_-)=(2,4)$. Without loss of generality, in the 2D-plots we have considered the same Gaussian distribution for the two anisotropy momenta by imposing $\bar{k}=\bar{k}_+=\bar{k}_-$ and $\bar{k}'=\bar{k}'_+=\bar{k}'_-$ in order to show the position of the probability peak (that occurs for $\bar{k}'=\bar{k}$) in a clearer way. 

\begin{figure}[h]
\includegraphics[width=0.95\linewidth]{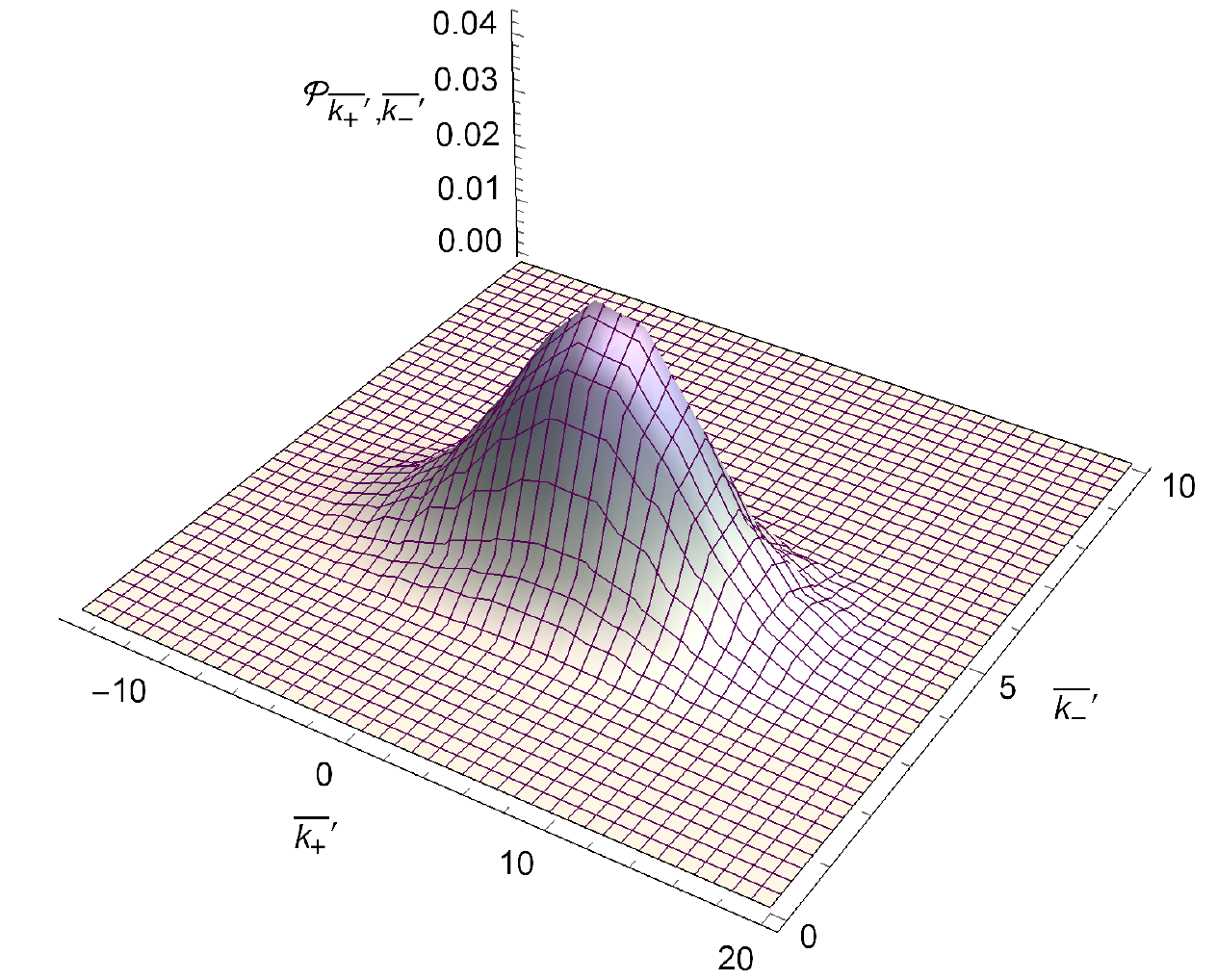}
\includegraphics[width=0.85\linewidth]{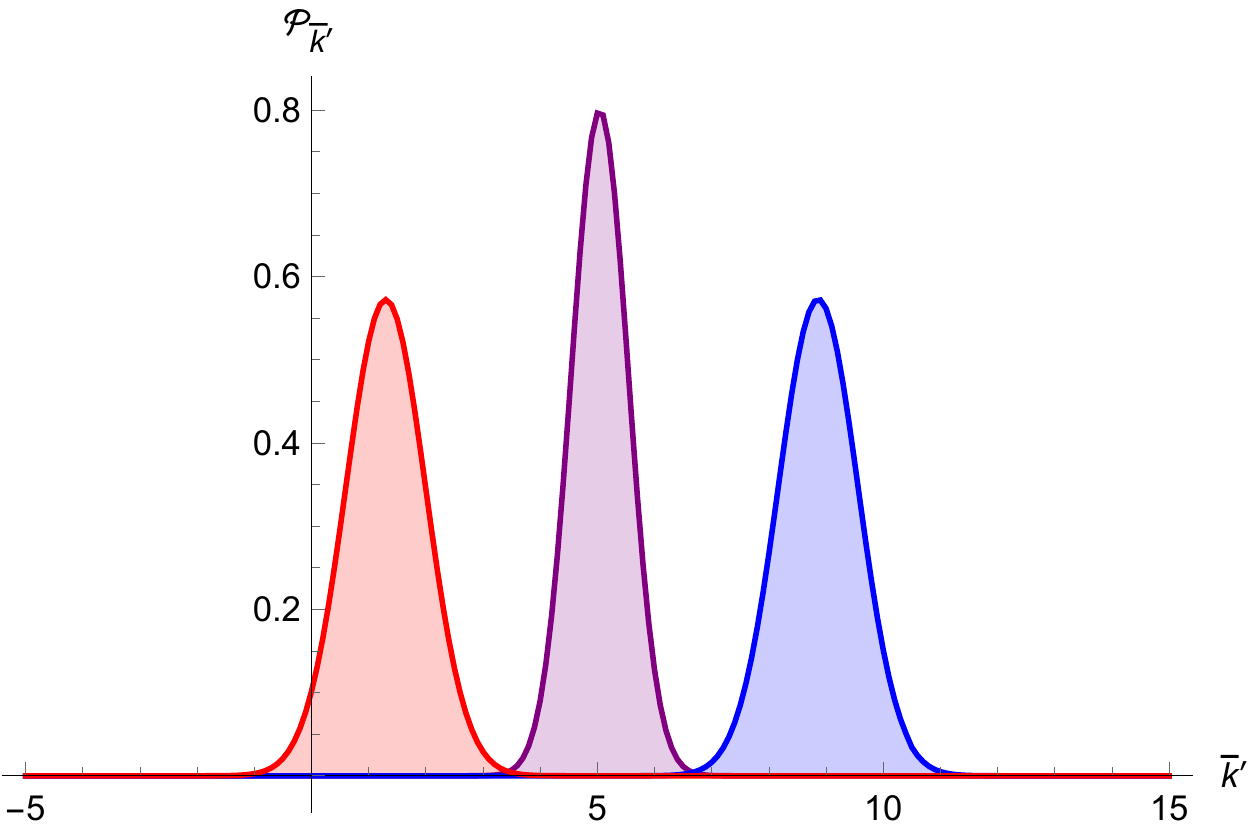}
	\caption{Top: 3D-plot of the normalized probability density $\mathcal{P}(\bar{k}'_+,\bar{k}'_-)$ of the Big Bounce transition for $\sigma_+=\sigma'_+=\sqrt{2\pi}$ and $\sigma_-=\sigma'_-=1/\sqrt{2\pi}$. We have considered $\lambda=1$, $\varepsilon=1/3$, $\bar{k}_+=2$, $\bar{k}_-=4$. Bottom: 2D-plots of the normalized probability density $\mathcal{P}(\bar{k}')$ of the Big Bounce transition for different values of the variances $\sigma_+=\sigma'_+$, $\sigma_-=\sigma'_-$. We have considered $\lambda=1$, $\varepsilon=1/3$, $\bar{k}_+=1$, $\bar{k}_-=9$ and $\sigma_+=\sigma'_+=1/\sqrt{2\pi}$, $\sigma_-=\sigma'_-=\sqrt{2\pi}$ in the red plot, $\sigma_+=\sigma'_+=1/\sqrt{2\pi}$, $\sigma_-=\sigma'_-=1/\sqrt{2\pi}$ in the purple one, $\sigma_+=\sigma'_+=1/\sqrt{2\pi}$, $\sigma_-=\sigma'_-=\sqrt{2\pi}$ in the blue one.}
\label{sigma}
\end{figure}

\begin{figure}[h!]
	\includegraphics[width=0.95\linewidth]{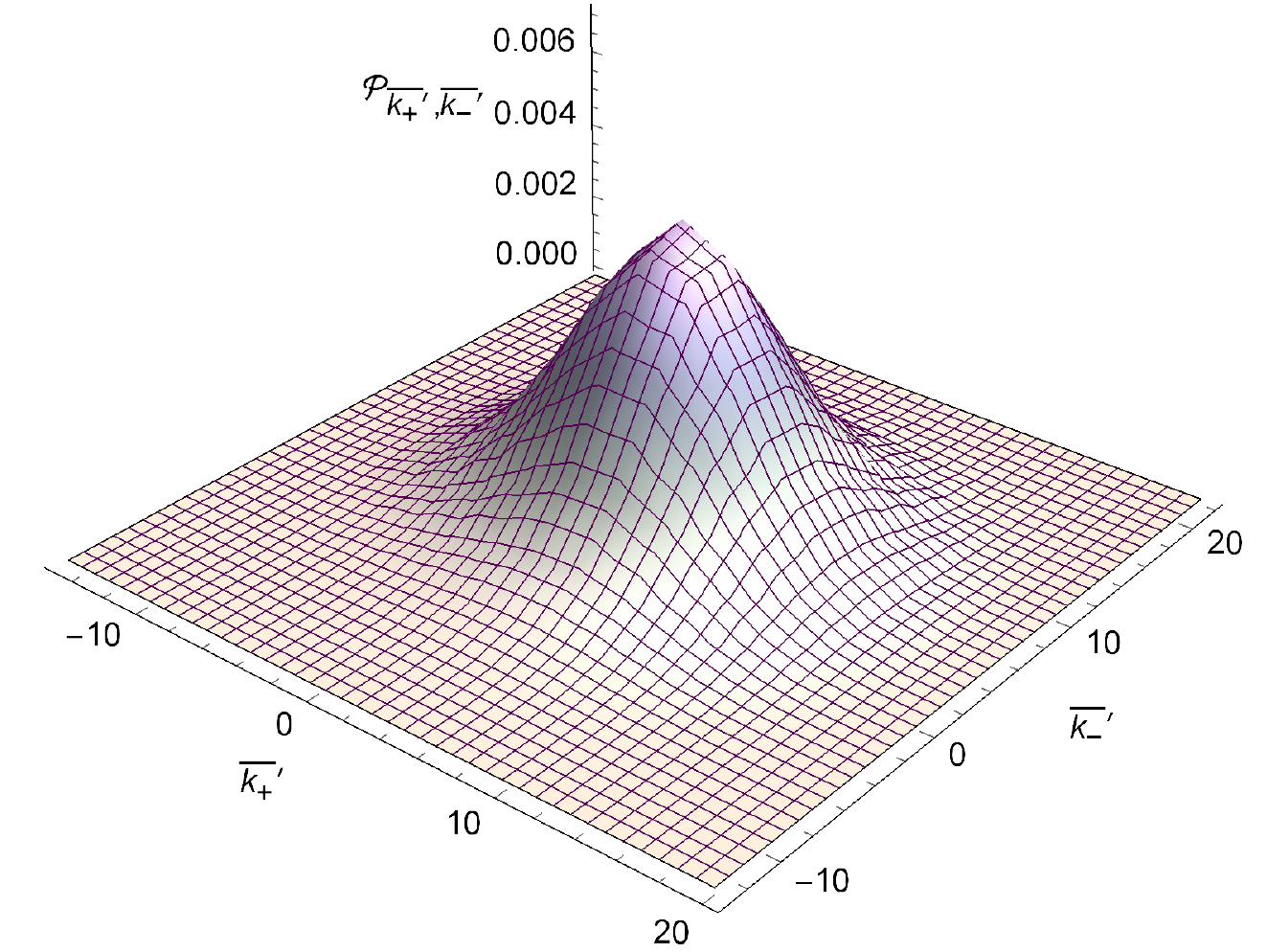}
	\includegraphics[width=0.85\linewidth]{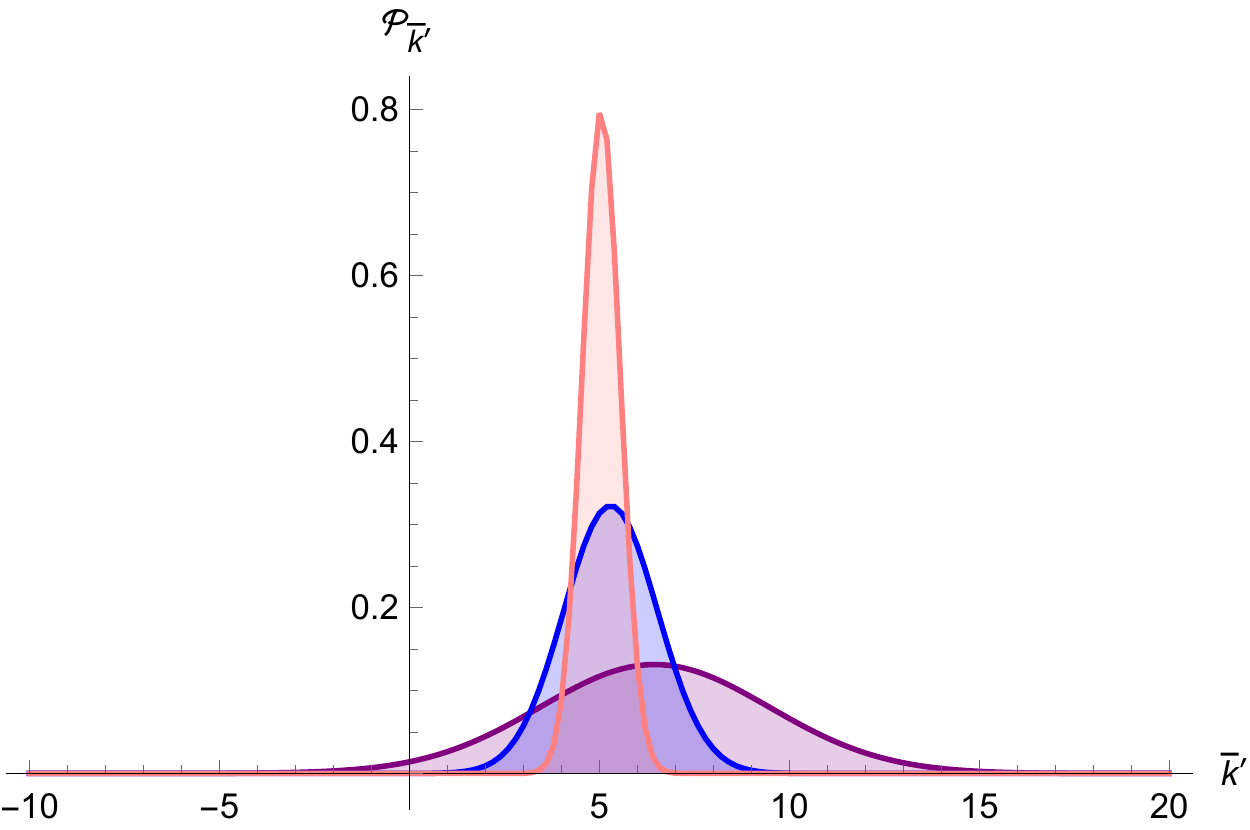}
\caption{Top: 3D-plot of the normalized
probability density $\mathcal{P}(\bar{k}'_+,\bar{k}'_-)$ of the Big Bounce transition for $\sigma_+=\sigma'_+=\sqrt{2\pi}$ and $\sigma_-=\sigma'_-=\sqrt{2\pi}$. We have considered $\lambda=1$, $\varepsilon=1/3$, $\bar{k}_+=2$, $\bar{k}_-=4$. Bottom: 2D-plots of the normalized probability density $\mathcal{P}(\bar{k}')$ of the Big Bounce transition for different values of the variance $\sigma$. We have considered $\lambda=1$, $\varepsilon=1/3$, $\bar{k}=5$ and $\sigma=1/\sqrt{2\pi}$ in the pink plot, $\sigma=1$ in the blue one,
$\sigma=\sqrt{2\pi}$ in the purple one.}
\label{sigma2}
\end{figure}

In Fig. \ref{sigma} we have only considered the hypothesis of $\sigma_+=\sigma'_+$ and $\sigma_-=\sigma'_-$, with the result that $\mathcal{P}(\bar{k}'_+,\bar{k}'_-)$ has a major variance along the direction in which the in-going packet is widely spread (see the 3D-plot) and also the position of the peak occurs no longer exactly in correspondence of the mean values $\bar{k}'_+$ and $\bar{k}'_-$ of the in-going wave packet. As a result, when the same mean values $\bar{k}'=\bar{k}'_+=\bar{k}'_-$ for the out-going wave packet are considered, the probability peak occurs exactly in correspondence of the average $\bar{k}'=(\bar{k}_++\bar{k}_-)/2$ only when the in-going wave packets are equally peaked, whereas when considering $\sigma_+\neq\sigma_-$ the peak occurs near the mean value of the more-localized in-going Gaussian distribution (see the 2D-plots). 
The relevant role of the wave packets variances  on the probability peak position is even more evident in Fig. \ref{sigma2}. In particular, in the 3D-plot we have considered all the Gaussian variances set to $\sigma=\sqrt{2\pi}$ we note that the peak of $\mathcal{P}(\bar{k}'_+,\bar{k}'_-)$ is slightly shifted with respect to the graph presented in Fig. \ref{prob}, where $\sigma=1/\sqrt{2\pi}$. Also, in the 2D-plots we can see that the peak of $\mathcal{P}(\bar{k}')$ occurs exactly in correspondence of $\bar{k}'=\bar{k}$ (here we have set $\bar{k}=5$ for the in-going Gaussian distributions) only when the in-going wave packet is highly peaked, whereas the bigger is the variance $\sigma$ the more is appreciable the shift of the peak with respect to the value of $\bar{k}$.  

In conclusion, we have demonstrated that the presence of the Big Bounce for the anisotropic Bianchi I Universe can be treated at a pure quantum level by means of a well-defined probability density, with different features depending on how the in-going and out-going wave packets are constructed. In other words, the presence of a potential that depends on the internal clock $\alpha$ makes possible to have a transition from a collapsing to an expanding wave function, before the Universe reaches the initial singularity, in a probabilistic sense. Indeed, the Universe wave function that emerges from the interaction with a time-dependent potential cannot be separated in its positive and negative frequency parts (except when the potential term becomes negligible). We remark that the transition we are considering is a fully quantum process, so the range of validity of the present procedure is restricted to a finite time region where the strong interaction potential is turned on. Clearly, when the Universe can no longer be described by a wave function, its dynamics follows the classical trajectories that derive from \eqref{eletc3}. Moreover, the probability amplitude tends to zero by construction when $\lambda\rightarrow0$, consistently with the fact that the transition of the Quantum Big Bounce is possible only in presence of a time-dependent potential able to create a mixed initial state for the collapsing Universe.

\section{Concluding remarks \label{C}}

In the analysis above we proposed the idea that a bouncing cosmology can emerge even in the framework of the WDW equation \cite{Misner69,MI,PC} (i.e. without introducing formalism and concepts of LQC \cite{Ashtekar2009}), as soon as the replacement of a collapsing Universe with an expanding one is viewed on a quantum level as a transition amplitude.

We analyzed a Bianchi I model in vacuum by identifying the Misner variable $\alpha$ as an internal time and basing our quantum study on a parallelism with the Klein-Gordon equation. We first emphasized the isomorphism existing
between the positive and negative frequency solutions and the expanding and collapsing Universe, respectively. Then, we have shown that the wave packets constructed to
approximate quasi-classically the evolution of the model unavoidably spread when $\alpha\rightarrow-\infty$ towards the cosmological singularity, coherently with the behavior of a $(2+1)$-dimensional massless relativistic particle. Thus, we could infer how the behavior of the Universe close enough to the initial Big Bang must be intrinsically a quantum phenomenon. Consequently, the absence of a quasi-classical trajectory which outlines a minimal volume is not sufficient to exclude the presence of a quantum transition from the collapsing to the expanding Universe, as soon as a physical mechanism able to break down the frequency separation is taken into account.

Therefore, we considered an ekpyrotic phase in the Kasner evolution, whose presence allows to deal with a non-zero probability that the collapse is replaced by an
expansion at a probabilistic level. In order to calculate this transition amplitude, having in mind to avoid any third quantization procedure, we adopted the standard method discussed in \cite{BjD}, which is applicable to any scattering process described in the formulation of the relativistic wave function. Thanks to the isomorphism between our model and a Klein-Gordon massless equation with a time-dependent potential, we estimated the probability amplitude associated to a collapsing wave packet that is projected onto an expanding one via the Green function.

We clearly demonstrated that the obtained probability density is well-defined and has a Gaussian shape. Its maximum occurs when the mean values of the quantities $k_+$ and $k_-$ (as
resulting from the Gaussian packets) essentially coincide in the collapsing and expanding Universe if and only if the corresponding wave packets are sufficiently localized. By increasing the variances, the position of the probability peak is shifted with respect to the case of high localization and occurs just nearby the mean values of the in-going wave packet. We propose this result as a notion of Quantum Big Bounce, which has also a quasi-classical symmetry in the collapsing and expanding branches when sufficiently localized wave packets (i.e. quasi-classical states) are considered. In other words, our analysis has the aim to give rise to the seed of a new scenario for the emergence of a bouncing cosmology, given the fact that the semiclassical notion of a modified dynamics, characterized by a finite maximum for the energy density, cannot be applied in many cosmological implementations of the quantum theory, since localized states naturally spread in quite general models when the singularity (or the presumed Bounce) is approached.

A further development is represented by the investigation of how the scenario proposed above can be implemented in the case of a generic relational time \cite{Rovelli}, i.e. at which extent we can make a direct identification of the frequency separation with the
nature of the Universe volume dynamics.
\nocite{*}
\bibliography{bib.bib}

\end{document}